\documentclass[10pt,conference]{IEEEtran}
\usepackage{amsmath,amssymb,amsfonts}
\usepackage{algorithmic}
\usepackage{graphicx}
\usepackage{textcomp}
\def\BibTeX{{\rm B\kern-.05em{\sc i\kern-.025em b}\kern-.08em
    T\kern-.1667em\lower.7ex\hbox{E}\kern-.125emX}}

\usepackage{comment}
\usepackage[dvipsnames]{xcolor}
\usepackage{hyperref}  
\hypersetup{colorlinks=true,urlcolor=blue,citecolor=blue,linkcolor=blue}   
\usepackage{outlines}
\usepackage{enumitem}        
\setlist{nosep}
\usepackage{makecell}

\usepackage[style=ieee,citestyle=numeric-comp,backend=biber]{biblatex}

\newboolean{redactswitch}
\setboolean{redactswitch}{false} 

\newcommand{\redact}[1]{\ifthenelse{
    \boolean{redactswitch}}{{[Redacted]}}{
    {#1}}}

\hypersetup{
           breaklinks=true,
}

\usepackage[hyphenbreaks]{breakurl}

\begin{document}

\title{
A Holistic Approach to Quantum Ethics Education\\ \vspace{0.5cm}
\large{The Quantum Ethics Project\\
\href{https://www.quantumethicsproject.org}{quantumethicsproject.org}}
}
\author{

\IEEEauthorblockN{Joan \'Etude Arrow}
\IEEEauthorblockA{\textit{Institute for Quantum Computing} \\
\textit{University of Waterloo}\\
Waterloo, ON, Canada \\
joan@quantumethicsproject.org}
\and
\IEEEauthorblockN{Sara E. Marsh}
\IEEEauthorblockA{\textit{Instructional Designer} \\
\textit{Ontario Certified Teacher (OCT)}\\
Kitchener, ON, Canada \\
smarsh@uwaterloo.ca}
\and
\IEEEauthorblockN{Josephine C. Meyer}
\IEEEauthorblockA{\textit{Department\ of Physics} \\
\textit{University of Colorado Boulder}\\
Boulder, CO, US \\
josephine.meyer@colorado.edu}
}

\maketitle

\begin{abstract}
This paper first provides an overview of the growing subfield of quantum ethics, including a working definition; research to date into social, economic, and political implications of quantum technologies; and directions for future research. Second, it introduces the Quantum Ethics Project (QEP), its activities to date, and its organizing philosophy. The third section reports on QEP’s ongoing curriculum development work, i.e. creating one of the first full-length courses on Ethics and Social Impacts of Quantum Technology. We outline the pedagogical approach being taken in the course design, including key learning outcomes, topic areas, teaching methods, and rationale. Finally, we discuss current limitations and future areas of attention, such as drawbacks to teaching ethical reasoning and ideas for assessment and implementation. 
\end{abstract}


\section{Introduction}

Rapid advancements have begun to cause a major shift in quantum information science, which has variously been described as ``moving from a purely scientific field to a technological one'' \cite{Williams:2023} and ``the world [being] on the cusp of a quantum revolution'' \cite{Myers:2023}. Billions in new investment is pouring in from public and private sources and interest in the field overall is growing rapidly. Faced with this veritable boom, researchers, technologists, and students are increasingly asking questions about what exactly quantum technology is promising to improve and for whom, what impacts it could have on people’s lives, who has the power to shape its development, what it might mean to do so “ethically” or “responsibly,” and how we can avoid repeating past mistakes (e.g.\ in AI/machine learning). The Quantum Ethics Project (QEP) is a grassroots effort founded in 2021 with the goal of bringing together all those who want to ask and answer these important questions. Since then, it has grown into a worldwide network and active online community of quantum enthusiasts who are exploring the intersection of quantum technology and society \cite{QEP:2023}. A key concern voiced by QEP network members is the lack of (a) instruction in this area in quantum degree programs which are training the next generation workforce and (b) resources that can bridge disciplinary divides and enable current technologists and researchers to participate in these critical debates. This is why one of QEP’s primary goals is to develop high-quality, open-access, modular educational materials for the quantum community that can be integrated into programs across educational levels and career stages.


\section{What is quantum ethics?}
Quantum ethics is an emerging academic discipline that draws heavy inspiration from discussions of ethics in the fields of Artificial Intelligence (AI) and Machine Learning (ML). In AI/ML ethics, researchers study the relationship between technological design and human impact. Individual human bias can collate in teams that are not diverse to produce technical designs which are themselves biased and have been shown to negatively impact marginalized groups. Quantum ethics discusses this relationship between technology and society in the context of quantum computing, communication, sensing, and other quantum technologies.
\newline
\newline
\textit{\textbf{Definition:}} Quantum Ethics is the field of study concerned with the social, economic, and political implications of quantum technologies.
\newline
\newline
When considering the social impact of advanced technologies, whether they be AI or quantum, both the positive and negative impacts are unlikely to be evenly distributed. In the case of AI, big data, and large-scale computing, technological benefits are overwhelmingly concentrated in the hands of a few wealthy individuals and corporations – while the harms are overwhelmingly concentrated in poor communities of color as well as women and gender-diverse people \cite{ONeil:2016,Eubanks:2018,Benjamin:2019, DIgnazio:2020}. Because of this inequity, the Quantum Ethics Project takes an equity-centered approach to quantum ethics.
We believe that quantum technological design cannot be ethical without also being equitable. Quantum technologists have a responsibility to learn from the mistakes of AI and to be proactive in anticipating the ways quantum may be misused to harm vulnerable communities. We must ensure quantum technology is designed and used for the \textbf{greatest and most equitable public good.}

\section{Background: Why quantum ethics? Why now?}

\subsection{Current context}

As the second quantum revolution matures, and as investment continues to grow, quantum technologists are beginning to increasingly use the term ‘when’ rather than ‘if’ quantum delivers an advantage and begins to impact society. Although the future of quantum remains uncertain, we need to start looking ahead and anticipating potential (mis)uses. This proactive approach learns from the mistakes of AI, as many believe AI ethics arrived too late to substantially influence the culture of design in the field \cite{Blackmore:2015,Hagendorff:2020,Inglesant:2021}. Once large revenues are being generated, efforts to regulate or otherwise reign in an industry that engages in profitable harm will be complicated by pushback from that industry, which has strong financial incentives to oppose safeguards that will hamper profits or require fundamental shifts to business practices. This highlights the urgency of bringing conversations about quantum ethics forward as soon as possible. Many quantum researchers and students desire that our research not produce technologies that hurt anyone, and this simple core value is what unites all quantum ethicists as we strive for a brighter quantum future. Starting conversations now about equitable quantum design \cite{Roberson:2023} -- being willing to ``own the unknown'' \cite{deJong:2022} -- is crucial to securing that future.

Over the past decade, Responsible Innovation \cite{Stilgoe:2013} has emerged as an academic framework for the responsible development of new technology, which includes four core principles of \emph{anticipation}, \emph{reflexivity}, \emph{inclusion} and \emph{responsiveness}. Approaches under this framework are very much rooted in ethics, seeking to foreground public goods and consider impacts on society. Along with considering how to practically apply concepts like accountability and liability, which tend to relate to the past, they place particular emphasis on forward-looking aspects of responsibility, such as \emph{taking care}.

Recent scholarship has adapted this field of study to quantum technologies \cite{Ingelsant:2018,Inglesant:2021,TenHolter:2021,TenHolter:2022, Kop:2023TowardsRQ, Kop:2023TenPrinciples}, making Responsible Innovation a particularly prominent theoretical framework for envisioning an ethical quantum revolution. Our work builds upon the principles of Responsible Innovation with the goal of bringing this powerful tool into the mainstream. At the same time, we recognize that discourse around Responsible Innovation has sometimes been employed in ways that privilege traditionally dominant perspectives \cite{Genus:2018}; whenever possible, our work at the QEP seeks to center the perspectives and epistemologies \cite{Prescod-Weinstein:2020} of those most often marginalized in both the study and the practice of science and technology to mitigate this tendency. 

\subsection{Quantum information technologies}
\label{sec:qistech}

At the core of this second quantum revolution is the shift from quantum-as-fundamental-physics to quantum-as-information-science: technologies that leverage the laws of quantum mechanics to circumvent classical limits on information processing. While the range of quantum information technologies is vast, at least three quantum information technologies have demonstrable possibility for significant market and societal disruption should they achieve commercialization, namely computers, communication networks, and sensors \cite{Pautasso:2021,Batra:2021}.

\subsubsection{Quantum computing}

While classical computers encode information as bits holding the value 0 or 1, quantum computers storing information in qubits (quantum bits) take advantage of the quantum properties of superposition and entanglement to perform calculations that are potentially intractable on a classical computer. 
Shor's quantum algorithm \cite{Shor:1994} can factor integers exponentially faster than the best known classical algorithms, potentially breaching today's internet security protocols. Quantum computers have also been hypothesized to hold potential for pharmaceutical discovery \cite{Bova:2021}, materials development \cite{Bauer:2020}, machine learning \cite{Biamonte:2017}, and finance \cite{Orus:2019, Bova:2021}.

\subsubsection{Quantum networking and communications}

Stephen Weisner first proposed that quantum mechanics could be used for ultrasecure communication immune to the threat of eavesdropping \cite{Wiesner:1983}. In 1984, Bennett and Brassard \cite{Bennett:1984} published a protocol for provably secure key distribution using a quantum channel, and many advances have followed. Quantum networks could find use in protecting sensitive data pertaining to trade secrets or national security. Recent work has demonstrated secure quantum key distribution from satellite to ground \cite{Liao:2017}, a key milestone toward global quantum networks.

\subsubsection{Quantum sensing}

Sensors exploiting the laws of quantum mechanics for ultraprecise measurements of time, gravity, electric field, and other quantities are rapidly approaching commercialization, with applications ranging from position, navigation, and timing (PNT) to mining and archaeology \cite{Kantsepolsky:2023}. Quantum sensors based on frequency comb spectroscopy have even shown promise in COVID breath detection \cite{Liang:2023}.

\subsection{Specific ethical issues in quantum technologies}

\subsubsection{Rhetoric and media hype}

There is growing concern in the quantum community about quantum media hype and its effect on society. Concerns have been raised about the impact of quantum hype on national security \cite{Smith:2020}, the business community, STEM education \cite{Meyer:2023TPT}, and the general public \cite{EPSRC:2018,Meinsma:2023}. There is also the risk of quantum hype creating an investment bubble \cite{Gibney:2019,Ezratty:2022}, with a resulting crash in research funding that could slow innovation and put numerous highly-educated researchers out of work, disproportionately harming those scientists from marginalized communities and the Global South that quantum diversity initiatives seek to benefit. The rhetoric used around quantum technologies matters too \cite{Palacios:2019}, at times privileging narratives of techno-nationalism over international cooperation \cite{EPSRC:2018,Roberson:2021QST,McKay:2022}. How can quantum researchers leverage our preeminent position in shaping public dialog \cite{Roberson:2021Minerva} to promote scientific honesty and social responsibility? 

\subsubsection{Privacy and cybersecurity}

Quantum technologies pose a double-edged sword for cybersecurity: while quantum networks promise an era of ultrasecure communications, quantum computers running Shor's algorithm might successfully break existing internet security protocols \cite{Mosca:2018} before existing classical channels migrate to quantum-safe protocols. How can we ensure a quantum-proof internet is built proactively rather than reactively and that individuals and businesses (especially those who lack the means to invest in quantum networks) are protected from the dual threats of government surveillance \cite{deWolfe:2017} and quantum cybercrime? What do quantum networks and sensors mean for the human right to privacy \cite{Krishnamurthy:2022}?

\subsubsection{Military applications and geopolitics}

Quantum technologies are being eyed for a number of prominent applications in defense, from navigation in GPS-denied environments to detection of underground bunkers \cite{Ingelsant:2018, Krelina:2021,Sayler:2021}. With both the US and Chinese militaries investing heavily in quantum research, some are beginning to worry these investments could trigger a Cold War-esque quantum arms race \cite{EPSRC:2018,Lele:2021}. All three quantum technologies we discuss in Sec.~\ref{sec:qistech} have both important military and civilian applications; civilian scientists are not immune to having their work applied to military purposes. What are the ethical obligations associated with developing these technologies, especially when powerful funders such as the US Department of Defense are involved? How can we ensure quantum technologies are used for peaceful purposes?

\subsubsection{Equitable distribution of benefits}

Quantum ethics also requires that the myriad benefits of new quantum technologies be distributed fairly \cite{TenHolter:2022}. With quantum investment remaining heavily concentrated in the Global North (and China), how can we prevent developing economies from being left behind? What happens if hedge funds bid up cloud quantum computing time to run portfolio-optimizing algorithms at the expense of less-resourced users \cite{EPSRC:2018} focused on education, medicine, or climate change? And without careful attention to diversity and inclusion, the quantum workforce risks replicating the diversity problems of its progenitor STEM fields, a concern amplified by the concentration of quantum educational programs at the most well-resourced universities \cite{Cervantes:2021,Perron:2021,Meyer:2023QST} and by restrictive visa regimes \cite{Malik:2022}. Under these circumstances, what does it mean for quantum technologies to be truly democratic \cite{Seskir:2023}?

\subsubsection{Sustainability}

Quantum computing may prove useful in managing variable electrical grids \cite{Giani:2021} and designing better electric vehicle batteries \cite{Rice:2021}, and quantum sensors may help to detect harmful methane leaks \cite{Crawford:2021}. Quantum computers might even one day achieve a ``green quantum advantage'' in energy usage compared to classical supercomputers \cite{Jaschke:2023}, but this analysis depends heavily on technical figures such as gate fidelity and ignores the likelihood that quantum energy advantage will be balanced by induced demand for intensive calculations. And on the whole, quantum technologies are not necessarily a win for the planet. For instance, is the use of quantum sensors to assist in the exploitation of the Alberta oil sands \cite{Sussman:2019} really as green as is claimed? What about sourcing exotic materials needed to build quantum computers in the first place \cite{McKay:2020}?

\section{About the Quantum Ethics Project}

\subsection{Our history}
The quantum ethics project was founded in September 2021 by Joan Arrow. While Joan was a Master’s student studying quantum algorithms and quantum machine learning, she felt that education and discussion around the ethics of quantum technologies was lacking in her program. From then on, she set out to meet with experts in quantum ethics and responsible innovation to better understand the potential societal implications of her research. A concrete goal of Joan’s early conversations was to create a full semester course on quantum ethics and to embed this course in the quantum information degree path at the Institute for Quantum Computing in Waterloo.

In 2022, after the first year of conversations, Joan had organized a growing online community on Discord, including assembling a team of three new collaborators to help map out the potential subjects and structure of a course on quantum ethics. This included quantum PhD students Rodrigo Araiza Bravo (Harvard U) and Darcy Morgan (UT Sydney), as well as education specialist Sara Marsh. This group conducted critical early work in mapping the field of quantum ethics and responsible innovation and laid the foundation for the QEP’s global leadership in quantum ethics education.

In early 2023, the QEP, having built a network of over 100 experts and enthusiasts, moved into its second phase of growth and activity. The leadership team expanded to include Sara Marsh, Rodrigo Araiza Bravo, and Perimeter Institute for Theoretical Physics graduate student Anna Knörr. In addition, the education team welcomed quantum education specialist Josephine Meyer and quantum networks expert David Sidi. The Quantum Education team at the QEP is developing quantum ethics workshops, seminars, and course materials for use by global academic institutions and industry partners.

\subsection{Pillars}
The QEP's efforts fall into three core pillars of activity:
\subsubsection{Education} The QEP education team develops and pilots quantum ethics curricula and course materials for audiences spanning the breadth of the quantum community, from students to quantum technologists and industry professionals.
\subsubsection{Research} Our research aims to jumpstart the new field of quantum ethics: What themes are shared with the ethics surrounding other advanced computing technologies such as AI? What topics are unique to quantum? By investing in fundamental research the Quantum Ethics Project is positioning itself as a global expert on the cutting edge of quantum policy, ethics, and responsible innovation. We invite others to join the conversation and help us integrate our findings into our educational materials to stimulate continuing discussions amongst students, researchers and the public worldwide.
\subsubsection{Diversity and outreach} Who is creating quantum technologies? The answer to this question will strongly influence what shape the quantum future will take. Hence, the QEP emphasizes creating opportunities for students from underrepresented backgrounds to get involved in quantum research and become a part of our community. We are also engaging with policymakers to ensure the emerging quantum workforce is composed of students from diverse sectors of society by investing in career opportunities and mentorship. Our hope is that our variety of backgrounds, ideas and concerns will contribute to making the quantum future equitable and just.

\section{Theoretical background: Ethics education for the second quantum revolution}



There is a growing sentiment in the quantum community that knowledge of quantum theory -- though important -- is insufficient by itself for developing a quantum-ready workforce. Tomorrow's workforce must exhibit competency beyond the physical theory, ranging from technician skills \cite{Hasanovic:2022} to non-technical ``soft'' skills such as teamwork \cite{Fox:2020,Aiello:2021}. 
What would it look like for quantum ethics education to be treated as equally foundational to tomorrow's quantum workforce? In this section, we unpack the theoretical and philosophical underpinnings of the Quantum Ethics Project's educational initiatives -- in other words, why we approach quantum ethics education the way we do, and what makes us unique.

\subsection{Ethics education as core part of the curriculum}

We at the Quantum Ethics Project are strong advocates for integrating ethics throughout the curriculum: If we wish to develop a well-rounded and socially-responsible quantum workforce, it is not enough to encourage our prospective physicists and engineers to take a few humanities electives. Ethics must be in the air we in quantum breathe -- and that means it must first become an integral part of our curriculum \cite{Hermeren:2013}. Quantum ethics education must be \textit{quantum-specific}, drawing on ethical issues and examples pertinent to the field and which today's quantum practitioners may one day encounter, and also \textit{recognized as a legitimate and essential aspect of training to be a quantum scientist} rather than a niche subfield. 
Disembodied ethics training that is not rooted in the real professional culture does not work \cite{Drake:2005,Hermeren:2013}; 
quantum ethics education must frame quantum ethics as part and parcel of being a quantum researcher so as to proactively shape the norms of the quantum community of practice \cite{Lave:1991} itself. 




\subsection{Macroethics and microethics}

Engineering ethicists classify professional ethics into two realms. \textit{Macroethics} emphasizes the quantum community's collective responsibility to society and issues best addressed at the community level, while \textit{microethics} focuses on the individual scientist's responsibility to disciplinary norms and conduct as part of the professional community \cite{Hudspith:1991}. Historically, engineering and professional ethics education has tended to focus on microethics at the expense of macroethics \cite{Herkert:2005}, though the ethics education community is beginning to shift toward pedagogies that integrate the two perspectives \cite{Canary:2012,Bielefeldt:2016,Knight:2016}. 

Our approach at the QEP integrates microethics and macroethics throughout the curriculum, using the concept of individual agency to link systemic macroethical issues with the daily decisions a scientist or engineer in the quantum industry may make. We believe that macroethical and microethical issues cannot be so easily separated in practice: individual practitioners often can and do have unexamined agency to address macroethical concerns, and individual microethical decisions are key building blocks of the quantum community culture that drives macroethical decision-making.

Case studies are a longstanding tradition in the science and engineering ethics education community \cite{Dyrud:2015}, and for good reason. However, case study approaches to ethics education have often been criticized for an overemphasis on microethics at the expense of macroethics, isolating scientists and engineers from the societal impacts of their work. 
We believe that careful case study design is key to avoiding this trap. One of our foundational case study worksheets, ``Climate urgency and the slowness of fault tolerance,'' centers on an individual researcher who is debating accepting funding from a climate-focused venture capitalist. 
The individual's choice whether to accept the funding (versus declining the award in favor of immediate decarbonization solutions needed to stave off the worst effects of climate change) is a microethical decision with substantial macroethical impacts.

\subsection{Holistic models of ethics education} 
In 2021, Clancy and Gammon \cite{Clancy:2021} made a compelling argument that the ultimate goal of ethics education ought to be the promotion of ethical behaviors. Indeed, we ought to be aware that ethics education without attention to subsequent behavior carries a potential moral risk: students might leverage sharpened ethical reasoning skills to rationalize or justify (rather than refrain from) morally-dubious behaviors, particularly in moral grey areas when careers, finances, or reputations are at stake. We briefly discuss two holistic frameworks centered on ethical behavior that guide our educational efforts.

\subsubsection{The four component model} Narvaez and Rest \cite{Narvaez:1995} identify four components of ethical decision-making: moral sensitivity (awareness of possible choices and their moral implications in social situations), moral judgment (effectively discerning the most moral choice), moral motivation (prioritization of moral values over other considerations), and moral implementation (possessing the skills and self-control to follow through on a moral decision). They argue that all four must be present for an individual to ultimately exhibit ethical behavior. 

Historically, ethics education has tended to focus primarily on moral judgment given that it is the component easiest to influence and measure \cite{Bebeau:2002}. The risk is that students become skilled at making ethical decisions when confronted with off-the-shelf case studies but fail to pair this with the self-awareness to recognize novel ethical issues and act according to one's principles. While of course no ethics intervention can guarantee ethical behavior, we intentionally design our curricular materials with all four components in mind.
For instance, our case study worksheet ``An international cybercrime ring'' features an ethical dilemma in which a board member at a quantum key distribution (QKD) company must balance two compelling ethical duties -- to protect the privacy of vulnerable users and to protect the public from cybercrime -- while also being mindful of one's fiduciary responsibilities to the company. This case study is powerful because it disrupts the ``wallet vs.\ heart'' mindset that so often underlies how ethical issues are framed, demonstrating that sometimes multiple selfless principles can themselves conflict.

More broadly, in our case study template, we challenge students to specifically consider all potentially conflicting stakeholders and issues (moral sensitivity), select a reasoned course of action amid no obvious right answer (moral judgment), critically analyze the harms of inaction (moral motivation), and identify barriers and incentives structures that may serve as obstacles to follow-through (moral implementation). 

\subsubsection{Virtue ethics} Aristotelian virtue ethics -- a philosophical theory grounded in the cultivation of a virtuous self -- provides a complementary perspective to the Four Component Model in that it views ethical behavior less as the culmination of a long moral reasoning process and more as a barometer of one's innate character. Traditionally, virtue ethics has primarily focused on the actions of lone individuals, but recent scholarship has begun to productively extend this theory to institutions and institutional actors as well \cite{Bright:2014,Cordell:2018}. Whereas traditional ethics curricula target acquisition of socio-cognitive ethical reasoning skills, character-centric models instead emphasize cultivation of virtues at the individual and structural level. These models recognize that ethics cannot be separated from professional practice: in the words of Schmidt (2014), ``Your practice \textit{is} your ethics!'' \cite{Schmidt:2014}. Recent work advocates incorporating virtue ethics as an additional philosophical perspective in engineering ethics education \cite{Schmidt:2014,Koehler:2020,McDonald:2022}, and we at the QEP seek to incorporate this counterperspective into our work. We are testing a virtue ethics module designed to be paired with any of our case study worksheets to enable educators interested in cultivating this perspective to expand upon it. The module features a variety of possible individual and institutional virtues discussed throughout the literature such as fairness, equity \cite{TenHolter:2022}, integrity \cite{McFarlane:2009}, and democratization \cite{Seskir:2023}. 

With its emphasis on self-responsibility, virtue ethics in particular demands an ability to reflect critically on one's own positionality amid an unequal society \cite{McCray:2019}. Our curricula aim to cultivate awareness of the perspectives of those historically excluded in quantum and in society, such as framing discussions of climate impact of quantum technologies in terms of the communities most directly impacted by climate change.

\subsection{Modularity and adaptability}

The needs of today's quantum workforce are rapidly evolving \cite{Fox:2020,Aiello:2021,Hasanovic:2022,Hughes:2022,Greinert:2023} and will likely continue to shift in unforeseeable ways as the quantum industry matures. Already there is an ongoing shift in the industry from a field dominated by physics Ph.D.'s to increasing numbers of positions where technician skills are most important \cite{Hasanovic:2022}. No longer restricted to Ph.D.-level specializations, quantum information science degree programs are proliferating at the undergraduate \cite{Perron:2021,Asfaw:2022,Dzurak:2022} and master's \cite{Plunkett:2020} levels and there has even been a push to bring quantum computing to high schools \cite{Satanassi:2021,Qtedu:2023,Q12:2023}. As quantum technologies become mainstream, there is also an increasingly-recognized need to develop a quantum-literate general public \cite{EPSRC:2018,Nita:2021} and a corresponding emphasis on the importance of outreach \cite{Faletic:2023}. Meanwhile, an increasing number of programs aim to bring underrepresented audiences into quantum using culturally-sensitive pedagogies \cite{Lee:2021,Porter:2022, GirlsInQuantum:2023,QubitByQubit:2023}.

In light of all of these converging trends, we recognize that a one-size-fits-all model of quantum ethics education can never be truly scalable. Quantum ethics education must be easily adaptable to a variety of audiences in terms of backgrounds, levels, and available course time -- a dedicated quantum ethics course is great, but so is a one-hour lecture on quantum ethics in an introduction to quantum computing course if  that's all that can be accommodated given scheduling constraints. In prior work, we interviewed quantum information science instructors to identify barriers and opportunities to incorporating quantum ethics education into the classroom \cite{Meyer:2022ASEE}. In particular, instructors requested modular resources that could be incorporated into a lecture or two of a quantum technologies course coupled with simple instructions for facilitation by non-experts \cite{Meyer:2022ASEE}. Our approach incorporates this feedback wherever possible, recognizing that retrofitting ethics into existing courses is one of the easiest ways to scale quantum ethics education in the near future. And our interactive, (semi)-non-hierarchical model of learning \cite{Bury:2018} -- in which the instructor, as facilitator, and students learn and challenge one another together through discussion -- is particularly well-suited for instructors who may lack experience with the subject of quantum ethics or with teaching subjects outside the paradigms of the hard sciences and engineering. As we move toward developing a full-semester pilot course (see Sec.~\ref{sec:full-semester}), we intend to weave many of the same resources we have developed for our shorter workshops into a broader whole.

\section{Educational initiatives at the QEP}
\subsection{Workshop at Perimeter Institute}

We collaborated with the Perimeter Institute for Theoretical Physics instructors Drs.\ Lauren Hayward and Eduardo Martin-Martinez to deliver an Introduction to Quantum Ethics Workshop for Master's students in the Perimeter Scholars International (PSI) program in March 2023. Drs.\ Martinez and Hayward were teaching Introduction to Quantum Information and Machine Learning for Many Body Physics respectively, and combined sections for our guest workshop. Students could attend either in-person or online. 

The workshop consisted of a 15-minute lecture followed by a 45-minute discussion. The lecture defined quantum ethics and gave an example case study on quantum machine learning entitled ``Quantum Computing for the 1\%''. The lecture was book-ended by two anonymous polls, which prompted participants for their initial and post-lecture reactions of what they associate with the term ``Quantum Ethics,'' which were then aggregated live into a word cloud. For the discussion, we separated students into small groups, each of which was provided with one of two ethical dilemmas (``Climate urgency and the slowness of fault tolerance'' and ``An international cybercrime ring'') to discuss. While students discussed in their groups, facilitators circulated to answer any questions that arose and provided further prompting based on the directions of students' conversations. Finally, the groups came together as a class to share their thoughts and conclusions, transitioning to a whole-class discussion about the common threads that emerged across different scenarios, and takeaways for thinking about ethics and social impact in quantum.


\subsection{Reading Group}
We are currently running a 10-week reading course that combines readings from the quantum ethics literature as well as readings from contemporary technology ethics and technology in society sources (e.g.~\cite{ONeil:2016,Eubanks:2018,Benjamin:2019,McKay:2020,Cerezo:2021, Coenen:2022, Rejeski:2022,Harris:2023,Seskir:2023}). In addition to the readings, a core group of 10 students (primarily undergraduates) are conducting a research project applying the above readings and their associated critical frameworks to algorithms taken from the quantum machine learning literature. This group combines students from the Perimeter Institute's recently created Quantum Ethics Working Group and the University of Waterloo's student club FemPhys, as well as graduate students from the Institute for Quantum Computing.

\subsection{Forthcoming: A 3-hour tutorial}

Our next major project is a 3-hour tutorial workshop to be debuted at the IEEE International Conference on Quantum Computing \& Engineering (QCE) in September 2023. The tutorial is intended to reach quantum professionals and practitioners at various levels of experience, and this implementation will be our first pilot outside the university setting. 
The tutorial will expand upon our workshop at Perimeter Institute, addressing the learning goals that attendees will be able to:
\begin{enumerate}
    \item Articulate the basic principles of quantum ethics and responsible innovation
    \item Identify and reason about certain specific ethical issues related to quantum technologies, including issues specific to their own work
    \item Critically interrogate claims about the merits and timescales of quantum technologies using the Gartner Hype Cycle \cite{Dedehayir:2016}; distinguish genuine near-term ethical issues from hype and fear-mongering
    \item Analyze how academic and corporate incentives structures, as well as a researcher's individual positionality, can serve to support or impede ethical decision-making (e.g. balancing fiduciary and societal responsibilities)
    \item Apply tools and principles of ethical decision-making to realistic ethical dilemmas similar to those they are likely to encounter in their employment or research
\end{enumerate}

As with other QEP initiatives, the tutorial will be interactive and include mini-lectures, large-group discussions, and small-group case studies and reflections. Table~\ref{tab:tutorial-outline} provides an agenda for the tutorial as an example of how QEP educational materials can be interwoven into a comprehensive curriculum.

\begin{table}[]
    \centering
    \begin{tabular}{c l}
        \hline \hline
        \textbf{Timing} & \textbf{Activity} \\ \Xhline{1pt} 
         0:00 & Intro and group norms \\
         0:10 & Interactive lecture: What is quantum ethics? Why now? \\
         0:35 & Discussion: Ethical issues in quantum technologies \\
         0:45 & Large-group guided case study: IBM Eagle \\
         1:05 & Discussion: Researcher positionality and incentives structures \\
         1:15 & Mini-lecture: Ethical frameworks \vspace{5pt} \\
         1:25 & BREAK \vspace{5pt} \\
         1:30 & Mini-discussion: Equity and considering all perspectives \\
         1:35 & Structured small-group case studies: Quantum ethics in action \\
         2:10 & Groups report back \\
         2:15 & Discussion: Ethical decision-making, lessons learned\\
         2:25 & Reflection activity: Quantum ethics in your research \\
         2:40 & Closing discussion: Exploring researcher agency \\
         2:55 & Wrap up and closing survey \\
        \hline \hline \\
         
    \end{tabular}
    \caption{\textup{Outline for our upcoming 3-hour workshop at IEEE QCE 2023. This example lesson plan demonstrates how specific QEP educational materials (lecture slides, discussion prompts, case study worksheets) can be combined into an integrated lesson plan.}}
    \label{tab:tutorial-outline}
\end{table}

\subsection{Full semester course}
\label{sec:full-semester}
A key impetus for creating the QEP was the lack of a university course or other accessible educational materials available on the ethics and social implications of quantum technologies. A scan of the landscape has made clear that no such full-length, permanent, core-topic course exists to date in North America, despite the explosion of graduate programs in quantum information and quantum technology \cite{Meyer:2023QST}. Thus, the QEP education team is developing such a course targeted towards or suitable to upper-year undergraduate and graduate students. Ultimately, we feel it is a necessary first step that every future quantum specialist should encounter this topic in their training before they enter the workforce. Our goal is to ensure every quantum information program in North America includes one of our courses on quantum ethics.

All of our activities to date have targeted this key objective: We have been steadily scaling up our content and materials, progressing from the first initial 1-hour guest lecture at Perimeter Institute, to the 3-hour tutorial coming at IEEE Quantum Week, and beyond. Our next major milestone involves drafting a syllabus for the full course and identifying a pilot site. After revision and refinement based on that first iteration, our aim is to make these materials available at other institutions and assist with integration into new and existing programs.

A key part of this work has been connecting with experts in STEM ethics education to make sure that our educational materials are theoretically grounded, utilize research-based teaching methodologies, and integrate robust evaluation and accountability mechanisms to ensure they are high quality and will actually meet the desired learning objectives \cite{Aiello:2021}.




\section{Evaluating and refining our work}

A concern expressed by some in the quantum community is that the rapid push to develop educational materials is coming at the expense of curriculum quality. As such, we feel it is important to prioritize accurate and detailed evaluation of our work to ensure our curricular materials and approaches are maximally effective. While our efforts to date have been primarily intended to build internal capacity within the Quantum Ethics Project team -- so opportunities to evaluate progress so far have been few -- as we pivot toward more outward-facing initiatives (culminating in a planned full-semester course) we will have a number of opportunities to quantitatively and qualitatively evaluate our work and refine accordingly. Pilot testing with data collection comprises the next phase of this work and will be the subject of further publications. 

So far, we have had one outward-facing educational initiative so far where we have been able to collect evaluative data. In our debut workshop in March 2023 at Perimeter Institute, we collected anonymous data on student responses to a Mentimeter poll at the end of class. While the data collected was purely anecdotal, student responses reflected thoughtful engagement with the material and identified tangible takeaways they can carry over into their work:

\begin{itemize}
    \item \textit{``Evaluate hype when thinking about quantum ethics.''}
    \item \textit{``It's often hard to identify minority voices but it's important to do so.''}
\end{itemize}
We were also struck by students' sense of moral agency after the workshop, and their apparent commitment to implementing principles of quantum ethics in their work moving forward -- a major goal of the QEP:

\begin{itemize}
    \item \textit{``Physicists have some role in these [ethical] decisions.''}
    \item \textit{``[Even if] someone else is technically responsible for a financial or ethical decision, there is often something you can do as an individual.''}
\end{itemize}

In future work, we intend to build on these informal methods of evaluation and design a formal feedback survey for participants at the end of our workshops. This survey will be debuted at our accepted tutorial at the IEEE Conference on Quantum Computing and Engineering (QCE) in fall 2023. If and when our planned full-semester course materializes, we also see value in augmenting fixed surveys with optional focus groups as sites for structured inquiry into the effects of quantum ethics education on factors such as ethical reasoning and motivation. Such investigation would also help us understand \textit{why} STEM ethics interventions have the impact they do, a question underexplored in the existing literature.

Aiello \textit{et al.} \cite{Aiello:2021} expressly advocate the use of validated assessments from the discipline-based education research community in designing and refining quantum curricula. Such assessments are carefully drafted and refined over a span of years to ensure their psychometric validity and reliability, enabling them to be used for comparisons across instructors and institutions. They are used in physics education research, for example, to reliably evaluate the effectiveness of curricular interventions on student conceptual reasoning \cite{Wilcox:2015} and even on beliefs about the nature of scientific knowledge \cite{Adams:2006}.

Promising options exist for quantitative analysis of students' ethical reasoning through validated instruments such as the Defining Issues Test (DIT-2) \cite{Rest:1999}, which measures the extent to which students' ethical judgments exhibit features of each of Kohlberg's \cite{Kohlberg:1958} three levels of moral development.\footnote{Kohlberg's three levels of moral development \cite{Kohlberg:1958} can be summarized as preconventional (focus on consequences and rewards to the self), conventional (focus on social and societal approval), and postconventional (focus on social contract and universal ethical values as opposed to rigid law-and-order). Kohlberg theorizes that human moral development necessarily proceeds through these three levels in order, though schema for multiple levels are in practice employed simultaneously. In using the DIT-2 to assess ethics education interventions, the assumption typically made is that increased use of postconventional thinking corresponds to positive moral and ethical growth.} Another option is the STEM-specific Engineering and Science Issues Test, or ESIT \cite{Borenstein:2009}. Both assessments have been used extensively in ethics education research to measure the effectiveness of educational interventions (e.g.\ Ref.~\cite{Drake:2005, Wilson:2011, Hamad:2013, 
LaPatin:2022}). It is important to note that these assessments, while useful, are only designed to measure ethical reasoning -- one component of the four component model. As such, while results from the DIT-2 and ESIT will be useful in helping us refine our curricular materials, we acknowledge that at best they are an incomplete measure \cite{Bebeau:2002}; quantitative methods alone will not substitute for qualitative evidence of holistic character development. Thankfully, there is also a growing body of literature on qualitative methods in ethics education evaluation \cite{Watts:2017} that we intend to draw upon to validate our work.

\section*{Acknowledgment}

\redact{We thank David Sidi, Anna Kn\"orr, and Rodrigo Araiza Bravo for valuable feedback on this paper and contribution to the QEP intellectual commons. Author J.C.M. acknowledges the support of the NSF GRFP and the Out to Innovate Career Development Fellowship. The QEP thanks our sponsors and partners: Center for Quantum Networks, Union of Concerned Scientists, Institute for Quantum Computing at University of Waterloo, and Perimeter Institute for Theoretical Physics.}

\printbibliography

\end{document}